# PROJECT RISK MANAGEMENT MODEL BASED ON PRINCE2 AND SCRUM FRAMEWORKS

Martin Tomanek and Jan Juricek

Department of Systems Analysis, University of Economics, Prague, Czech Republic

## ABSTRACT

*There is a lack of formal risk management techniques in agile software development methods Scrum. The need to manage risks in agile project management is also identified by various authors. Authors conducted a survey to find out the current practices in agile project management. Furthermore authors discus the new integration framework of Scrum and PRINCE2 with focus on risk management. Enrichment of Scrum with selected practices from the heavy-weight project management methodology PRINCE2 promises better results in delivering software products especially in global development projects.*

## 1. INTRODUCTION

Agile methods grew out of the real-life project experiences of leading software professionals who had experienced the challenges and limitations of traditional waterfall development methodologies on projects after projects. The agile development frameworks are widely used and they don't contain any risk management techniques because it is believed that short iterative development cycles will minimize any unpredictable impact related to product development [1], [2]. However in larger projects or during development of complex products, especially in the global environment, the need of proper risk management is required. From the audit perspective, there is the clear control requirement "BAI01.10 Manage programme and project risk" defined by COBIT 5 that requires that project risks should be systematically identified, analysed, responded to, monitored and controlled. Additionally the risks should be centrally recorded [3, p. 125]. Additionally, controlling risk in software projects is considered to be a major contributor to project success [4].

The need to manage risks in agile project management is also identified by various authors. The SOA principles from the agile project management perspective were used to create a framework for understanding agile risk management strategies for global IT projects [5]. Main risk models and frameworks used by software engineers are discussed with conclusion that the risk management steps are required for delivery of quality software [6], [7]. Agile methodologies don't cover the risk management knowledge area that can be taken from project management frameworks like PMBOK [8]. Risks related to global software development projects using Scrum have been researched and a conceptual framework to mitigate them designed [9]. Also the increasing variety of security threats should be managed as risks in the agile development projects [10], [11].

The authors of this paper prove that this lack of risk management techniques in agile development can be fixed by aligning risk management techniques between project management framework PRINCE2 and agile product development framework Scrum. However the traditional risk management is heavily centred on documentation and agile development principles tries to avoid any unnecessary documentation and focus more on people iteration [12].

## 2. PRINCE2 RISK MANAGEMENT PROCESS

Prince2 provides a disciplined environment for implementation of risk responses based on identifying and assessing project risks. Lack of the proper risk management is the one of the leading factor, why the projects failed [13]. Risk management within PRINCE2 methodology contains three dimensions: risk management strategy (how risk management will be embedded in the project management activities, what is the risk tolerance and when is the exception triggered); Risk register as a tool for capturing and maintaining information of identified threads and opportunities (Project Support will typically maintain the Risk Register on behalf of the Project Manager); and a risk management procedure [14].

The project risk management process recommends 5 steps: identify, assess (estimation and evaluation), plan, implement and communicate. The first four steps are sequential, with the 'Communicate' step running in parallel; As well as techniques for identification, risk estimation and evaluation techniques and thread responses (avoid, reduce, fallback, transfer, share, accept). In the implement step, the risk owner and the risk actionee roles are defined. The risk owner is a named individual who is responsible for the management, monitoring and control of a risk, while the risk actionee is an individual assigned to carry out a risk response action or actions to respond to a particular risk or a set of risks. They support and take direction from the risk owner.

## 3. RISK MANAGEMENT IN SCRUM

Scrum doesn't define the formal risk management process or even the risk owner. However every Scrum artefact or meeting potentially help to identify or mitigate risks. The list of artefacts and meetings with related risks is described for example in [15]. If the Scrum master, the product owner or the development team want to manage risks in a more formal or in a more proactive way then they can use for example a simple risk register [1] or a risk burn-down chart [16]. If they do so then it will be recommended to prioritize the highest-value and highest-risk requirements first in the upcoming sprint [15].

In the Scrum guide [17] there is a concept of impediments. The impediment can be anything that keeps a team from being productive. From risk management perspective the impediment is equal to an issue (a materialized risk). The Scrum master is responsible to solve these impediments and let the development team work in effective environment.

The other risk management model, mentioned in [18], combines the PMBOK approach with Scrum. The author suggests using a risk board with two kinds of notes. The red notes are used to describe the risks and the yellow ones to describe the risk responses. Another good example how risks can be potentially mitigated in Scrum (such an intrinsic schedule flaw, specification breakdown, scope creep, and personnel loss and productivity variance) is mentioned by [19].

Some techniques have been developed to support the story prioritization based on inherit risks, for example the story-risk prioritization matrix [20] or the risk burn-down technique [21].

## 4. SURVEY

The authors of this paper conducted in April and May 2014 a survey with objective to found out how project managers manage risks in agile software development projects. The completed survey was received from 64 project managers working in Czech Republic and Slovakia on global software development projects.

The results of the survey:

- 67 % of respondents review the project risks on weekly basis.
- 87 % of respondents stated that the project manager is formally responsible for managing risks.
- 27 % of respondents stated that also the product owner is formally responsible for managing risks.
- Respondents were asked who primarily identifies the risks. The project team members (80%), the project manager (67 %), the development team member (40%) and the scrum master (40%).
- 73 % of respondents use a risk log or risk register to document and maintain risks. The rest of respondents 27 % use different techniques or don't document risks at all.

Respondents were also asked what risk attributes they document and keep watching. The Figure 1 shows the result. The most important risk attribute is the risk exposure (severity of impact) and likelihood that are defined during risk assessment and analysis exercise. The mitigation actions and responsible persons are the key elements of risk management. The responsible person executes the mitigation actions in order to mitigate the risks.

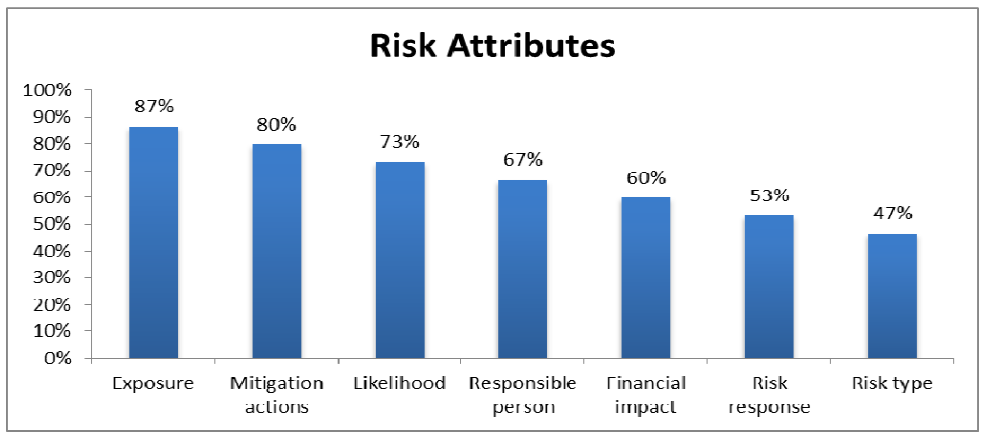

Figure 1. Risk Attributes. Source: The survey conducted by authors.

The Figure 2 shows the Scrum meetings where the project managers discuss the risks with the development team. The most critical meeting for identifying risks and planning mitigation actions is the sprint planning meeting. The review of existing risks and identification of potential risks introduced by the product increment is done in the spring review meeting. As the project manager needs to regularly review the risk status, the weekly status meeting is ideal for him or her. The weekly status meeting can be an extended version of the daily Scrum meeting.

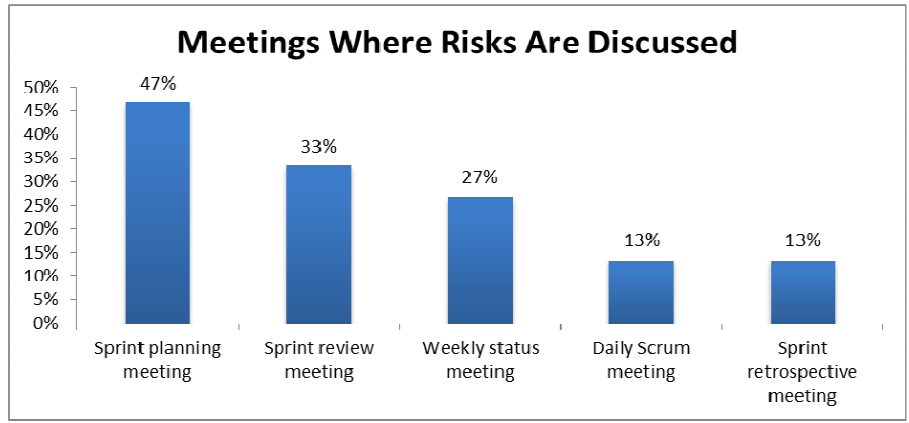

Figure 2. Risk Meetings. Source: The survey conducted by authors.

The last question was formulated as "Do you think that agile development frameworks should be enriched by the risk management techniques from the project management framework?" The majority of respondents (60%) think that the enrichment would be helpful and 40% think that there is no value in this enrichment.

## 5. TAILORING OF PRINCE2 PROJECT AND RISK MANAGEMENT TO MEET THE SCRUM APPROACH

At first, integrated process model should be created for tailoring the PRINCE2 project management process model to meet the Scrum approach. This conceptual model has been developed and it is described in [22]. In this paper this model was further reviewed, slightly modified and key meetings (ceremonies) and documents highlighted in red, see the Figure 3.

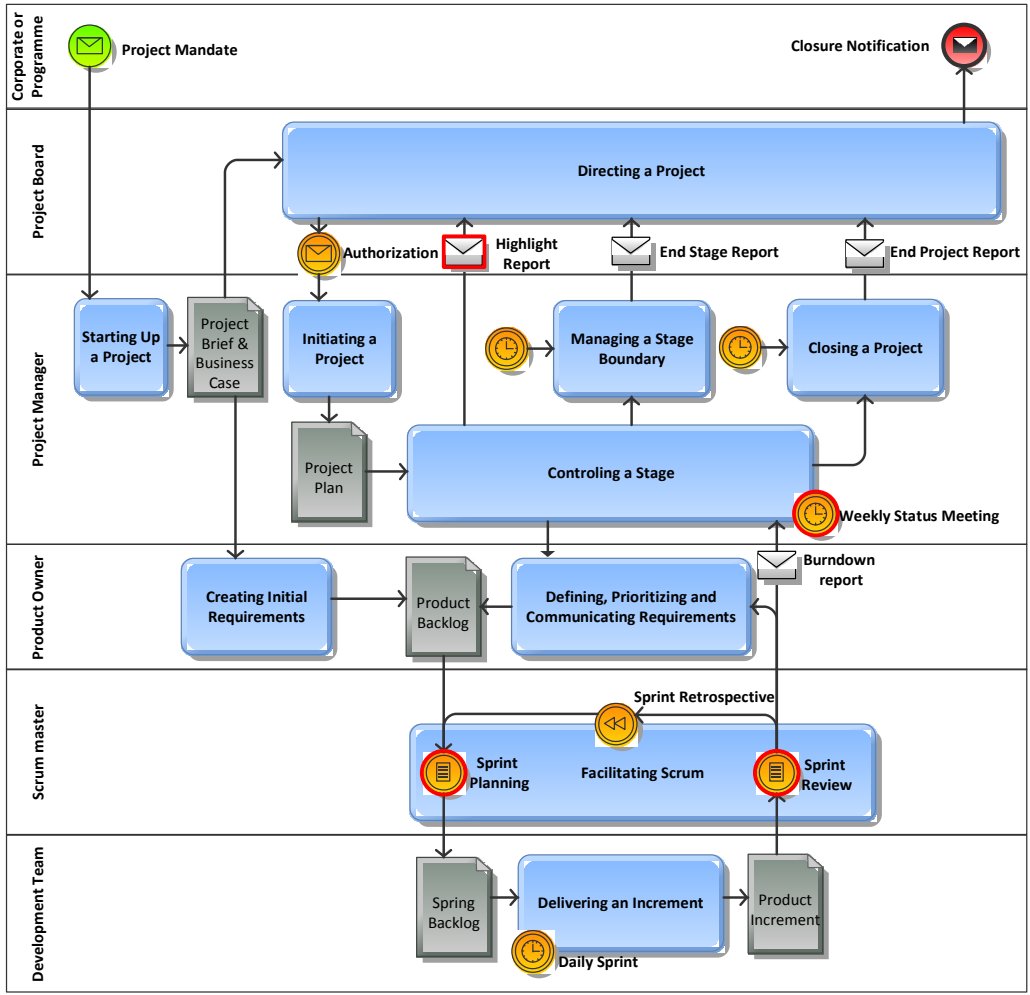

Figure 3. Process model integrating PRINCE2 and Scrum frameworks with highlighted risk meetings.
Source: Authors.

## 5.1. Key meetings and documents

The **sprint planning meeting** is the key meeting where the risks are identified. These risks are mostly related to the stories/software features that will be developed in the coming sprint. Based on the risk-story prioritization technique [20], the high-risk stories should be selected first and developed first. All the risks related to the developing features should be adequately assessed and evaluated by the development team. The risk attributes that are mostly used are: risk exposure, mitigation actions, likelihood, responsible person and financial impact (based on authors' survey).

The **weekly status meeting** should contain the review of the current sprint risks. This meeting should be led by the project manager. The Scrum master should also participate in this meeting because he/she has the overview of the risks and other impediments the development team is facing to. It is a good practise to invite also risk actionees who can comment on progress for the critical risks or risks with the lowest risk proximity. As a suitable technique for reviewing the total sprint risk exposure can be the risk burndown chart [21].

International Journal of Software Engineering & Applications (IJSEA), Vol.6, No.1, January 2015The **sprint review meeting** is focused on presenting the product increment to the product owner and other customers/key users. The risks related to the development of product features should be already mitigated and closed. If some risks are not fully mitigated then the residual risks should be presented here and accepted. Also new risks that were identified during the development and are not fully mitigated can be presented here and mitigation actions planned.

The **highlight report** is used by the project manager to inform the project board about the project progress. Information about the key risks should be part of this report.

The Table 1 describes the mandatory or optional risk related tasks that are done during the Scrum meetings. This does not mean that risk identification, assessment and responses would only be done during these meetings. Most of the work will be done between the meetings in the process "Delivering the product increment" that is represented by the Sprint. M means mandatory and it should be done every time or on a continuous basis, O means optional and it should be done only if needed.

Table 1. Risk Management Activities in Scrum Meetings and the sprint, Source: Authors.

| Scrum meetings | Risk Actions | | | | |
|---|---|---|---|---|---|
| | Identification | Assessment | Response Plan | Apply Response | Risk Approval |
| Sprint Planning | M | M | M | | O |
| Daily Sprint | O | O | O | | |
| Sprint Review | O | O | O | | M |
| Sprint | | | | M | |
| Retrospective | O | O | O | | |

## 5.2. Risk Escalation

One of the PRINCE2 principles is the Manage by exception. The project manager manages the project and tries to keep the project within its budget, deliver the project on time and in required quality. For these aspects the project manager has defined boundaries within which he or she can act. If these limits are exceeded then project board involvement is required. Additional boundaries are specified by the project manager for the development team. The development team also acts in defined limits and when these limits are exceeded then it must be escalated to the project manager. These limits can be set for example to: minimum number of stories/features to be delivered in the sprint, acceptable deviation for the sprint progress based on the burndown chart and also for the criticality of the identified risks.

## 6. CONCLUSIONS

Risk management in agile software development projects has been discussed by many researchers and various techniques have been proposed. However, based on the literature review executed by authors, there is no comprehensive model for embedding the project risk management process into the Scrum framework. Authors also executed a survey with objective to understand how project managers manage risks in the agile software development projects. As a final result of this paper the authors proposed to extend the conceptual framework for managing agile software development projects, based on Scrum and PRINCE2, by risk management aspects. The conceptual framework is extended by mandatory or optional risk related tasks that are done during the Scrum meetings. The major Scrum meetings and their relations to risk management activities have been discussed. The benefits of agile – scrum risk management include improving capacity to manage project uncertainties on the product delivery level and enhance communication of risks within the entire project organization.

## AUTHORS

**Martin Tomanek** graduated from applied informatics at the Faculty of Informatics and Statistics, University of Economics, Prague. Currently, he is PhD student at the Department of Systems Analysis, Faculty of Informatics and Statistics, University of Economics, Prague, where he develops the integrated framework based on PRINCE2, Scrum and other best practices used in SW development area.

**Jan Juricek** graduated from applied informatics at the Faculty of Informatics and Statistics, University of Economics, Prague. Currently, he is PhD student at the Department of Systems Analysis, Faculty of Informatics and Statistics, University of Economics, Prague, where he deals with agile principles, objectives and benefits in project management.